\documentclass[journal]{IEEEtran} 
\usepackage{amsmath,amssymb}
\usepackage{subfigure}
\usepackage{graphicx,graphics,color,psfrag}
\usepackage{cite,balance}
\usepackage{caption}
\allowdisplaybreaks
\usepackage{algorithm}
\usepackage{algorithmic}
\usepackage{accents}
\usepackage{amsthm}
\usepackage{bm}
\usepackage{url}
\usepackage[english]{babel}
\usepackage{multirow}
\usepackage{enumerate}
\usepackage{cases}
\usepackage{stfloats}
\usepackage{dsfont}
\usepackage{color,soul}
\usepackage{amsfonts}
\usepackage{cite,graphicx,amsmath,amssymb}
\usepackage{subfigure}
\usepackage{fancyhdr}
\usepackage{hhline}
\usepackage{graphicx,graphics}
\usepackage{subfigure}
\usepackage{array,color}
\usepackage{mathtools}
\usepackage{amsmath}

\newtheorem{definition}{\emph{\underline{Definition}}}

\newtheorem{lemma}{\emph{\underline{Lemma}}}
\newtheorem{corollary}{\emph{\underline{Corollary}}}

\newtheorem{proposition}{\emph{\underline{Proposition}}}

\newtheorem{remark}{\bf \emph{\underline{Remark}}}

\def\l{\left}
\def\r{\right}
\def\({\left(}
\def\){\right)}

\def\b0{{\mathbf{0}}}

\newcommand{\nn}{\nonumber}

\usepackage[top=0.5in, bottom=0.45in, left=0.5in, right=0.5in]{geometry}

\begin{document}

\captionsetup[figure]{name={Fig.}}

\title{Active-IRS Aided Wireless Network: System Modeling and Performance Analysis}
\author{Yunli Li, 
        Changsheng You, \IEEEmembership{Member, IEEE}, 
        and
        Young Jin Chun, \IEEEmembership{Member, IEEE}

\thanks{Y. Li and Y. J. Chun are with City University of Hong Kong (e-mail: yunlili2-c@my.cityu.edu.hk; yjchun@cityu.edu.hk). C. You is with Southern University of Science and Technology (SUSTech) (e-mail: youcs@sustech.edu.cn).}}


\maketitle

\begin{abstract}

Active intelligent reflecting surface (IRS) enables flexible signal reflection control with \emph{power amplification}, thus effectively compensating the product-distance path-loss in conventional passive-IRS aided systems. 
In this letter, we characterize the communication performance of an active-IRS aided single-cell wireless network. To this end, we first propose a \emph{customized} IRS deployment strategy, where the active IRSs are uniformly deployed within a ring concentric with the cell to serve the users far from the base station. Next, given the Nakagami-$m$ fading channel, we characterize the cascaded active-IRS channel by using the \emph{mixture Gamma distribution} approximation and derive a closed-form expression for the mean signal-to-noise ratio (SNR) at the user averaged over channel fading. Moreover, we numerically show that to maximize the system performance, it is necessary to choose a proper active-IRS density given a fixed number of total reflecting elements, which significantly differs from the passive-IRS case for which the centralized IRS deployment scheme is better. Furthermore, the active-IRS aided wireless network achieves higher spatial throughput than the passive-IRS counterpart when the total number of reflecting elements is small.
\end{abstract}
%
%
%
%
%
\section{Introduction}
Intelligent reflecting surface (IRS) has emerged as a promising technology to enable a reconfigurable radio propagation environment by smartly controlling the signal reflection at its reflecting elements. Moreover, IRS operates at full-duplex mode \cite{ye2021spatially} and is flexible to be deployed in the environment, which thus has attracted extensive attention in recent years to incorporate IRSs into traditional wireless networks to enhance the communication performance \cite{wu2021intelligent}.

Most of the existing literature on IRS has considered the passive IRS, where the IRS can reflect signals only without signal processing/amplification capability. Besides the designs of IRS passive beamforming and channel estimation (see, e.g., \cite{9133142}, \cite{zheng2022survey}, \cite{sharma2020intelligent}), substantial research has also been devoted to analyzing the communication performance of passive-IRS aided wireless networks (see, e.g., \cite{lyu2020spatial, lyu2021hybrid}). 
For example, the authors in \cite{lyu2020spatial} analyzed the spatial throughput of a passive-IRS aided single-cell wireless network by applying the Gaussian distribution approximation for the cascaded channel and using the moment matching method to approximate the mixture received signal power as a Gamma distribution. This work was further extended in \cite{lyu2021hybrid} to study the performance of multi-cell wireless networks. However, the existing works focused on Rayleigh fading and moment matching modeling of the channels, which ignored the complexity of the real propagation environment and scarified the accuracy of channel modeling. Moreover, the passive-IRS aided network suffers severe product-distance path-loss from signal reflection, leading to limited communication performance enhancement, especially when direct line-of-sight (LOS) links exist. 

To address the above issue, another type of IRS, called \emph{active IRS}, has been recently proposed in the literature \cite{zhang2021active,long2021active}. Specifically, the active IRS is equipped with a reflection-type amplifier to enable simultaneous signal reflection and amplification, thus effectively compensating the product-distance path-loss. Moreover,  the authors in \cite{you2021wireless} show that for maximizing the achievable rate (AR), the active IRS should be deployed closer to the receiver with decreasing amplification power. 
Despite the above link-level analysis, the network-level analysis for active IRS is still lacking in the literature, which is particularly important for network planning and performance evaluation.

Motivated by the above, we study the communication performance of an active-IRS aided single-cell wireless network in this letter. Specifically, we first propose a \emph{customized} IRS deployment strategy, where the active IRSs are uniformly deployed within a ring concentric with the cell to serve the users far from the base station (BS). Next, under the Nakagami-$m$ fading channel model for each link, we characterize the cascaded active-IRS channel by the \emph{mixture Gamma distribution} approximation, which achieves higher accuracy than the widely-used moment matching method when the number of active reflecting elements of each IRS, $N$, is small. Then, we derive a closed-form expression for the mean signal-to-noise ratio (SNR) at the user averaged over channel fading and show that the received SNR linearly scales with $N$ and increases quickly in the low amplification power region. Last, simulations are presented to evaluate the average SNR and spatial throughput of the considered active-IRS aided wireless network. It is shown that it is necessary to choose a proper IRS density given a fixed number of total reflecting elements to maximize the spatial throughput, which significantly differs from the passive-IRS case for which the centralized IRS deployment scheme is optimal. Moreover, the active-IRS aided wireless network achieves higher spatial throughput than the passive-IRS counterpart when the total number of reflecting elements is small.

\vspace{-0.3cm}
\section{System Model} \label{sectionSystemM}

\begin{figure} [t]   
	\centering
	\includegraphics[height=2cm,width=7cm]{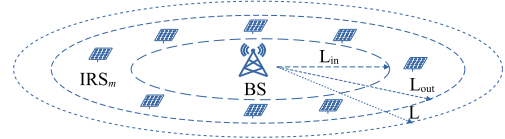}
	\caption{An active-IRS aided single-cell multiuser wireless network.}
	\label{fig:systemModel}
\end{figure}
We consider an active-IRS aided single-cell multiuser wireless system as shown in Fig.~\ref{fig:systemModel}, where multiple active IRSs are deployed in the network to assist the communication from one single-antenna BS to a group of $K$ single-antenna user equipment (UEs).\footnote{We consider the downlink communication for the single-antenna BS in this work, while the obtained results can be extended to the case of uplink communication as well as the multi-antenna BS by designing the BS's active beamforming properly.} We assume that the UEs are uniformly and randomly distributed in a disc cell area with a radius of $L$ meters (m), and the BS is centered at the origin. Moreover, similar to \cite{lyu2020spatial,lyu2021hybrid}, for ease of analysis, we assume that the bandwidth and each time slot are equally divided into $K$ orthogonal resource blocks, each randomly assigned to one UE, over which the channel is assumed to remain static. The current work can be extended to the case where the BS simultaneously communicates with multiple users at the same time, while the interference analysis is more complicated and thus left for future work.

\subsection{Active IRS Deployment and UE Association}
For ease of analysis, we introduce a customized and low-complexity IRS deployment strategy with nearest association policy to achieve a near-optimal performance with tractability.
We consider that $M$ active IRSs, each equipped with $N$ reflecting elements, are deployed to assist the downlink communication from the BS to the UEs. Unlike most existing works on passive IRSs that considered IRSs uniformly distributed in the whole cell, we assume a \emph{customized} IRS deployment scheme for the active-IRS aided wireless network. Specifically, the active IRSs are distributed in a ring concentric with the cell with an inner ring radius of $L_{\rm in}$ and an outer ring radius of $L_{\rm out}$ (see Fig.~\ref{fig:systemModel}). The rationale for this new IRS deployment is explained as follows. First, it was shown in \cite{you2021wireless} that to encourage the active IRS to operate at the power amplification mode, the active IRS should be deployed away from the transmitter by a certain range; thus, we consider the inner range $L_{\rm in}$. Moreover, for UEs in close proximity to the BS, the BS$\to$UE link is sufficiently strong in general;  thus, it is more beneficial to deploy IRSs for assisting UEs far from the BS \cite{zeng2022throughput}. Second, unlike the passive-IRS case for which IRS is preferred to be deployed near the BS/users for reducing the product-distance path-loss, it was shown in \cite{you2021wireless} that the active IRS should be deployed between the BS and UEs to achieve favorable rate performance; thus we consider the outer range $L_{\rm out}$ accounting for the cell-edge UEs.

Under the above customized active IRS deployment strategy, we consider the following nearest UE association policy for ease of analysis: if the UE-BS distance is smaller than $L_{\rm in}$, the UE is directly associated to the BS; otherwise, it is associated to its nearest active IRS, for which we assume that the direct link is negligible due to long distance. Note that the customized deployment strategy with nearest-IRS association policy not only allows for tractable performance analysis but also achieves close performance to the uniformly deployment strategy with optimal-IRS association scheme that associates each UE to its best active IRS, as validated in Section \ref{Section: numerical results} through simulations \footnote{The optimal association policy with customized deployment strategy  may provide better performance, and this will be studied in future work.}.

\subsection{Channel Model and IRS Reflection Model}

Consider a typical UE and its nearest active IRS. Let $h_{\rm BU}=\sqrt{\zeta_{\rm BU}} g_{\rm BU}$ denote the channel from the BS to UE, where $\zeta_{\rm BU}\triangleq \epsilon d_{\rm BU}^{-\alpha_{\rm BU}}$ denotes the BS-UE path-loss with $\epsilon$ representing the reference channel power gain at a distance of $1$ m,  $d_{\rm BU}$ being  the BS-UE distance, and $\alpha_{\rm BU}$ being the corresponding path-loss exponent. Moreover, $g_{\rm BU}$ denotes the small-scale fading channel, whose amplitude is assumed to follow the Nakagami-$m$ distribution with parameter $m_{\rm BU}$.
Similarly, the BS$\to$IRS and IRS$\to$UE channels, denoted by ${\bf h}_{{\rm BI}}\in\mathbb{C}^{N\times 1}$ and ${\bf h}^H_{{\rm IU}}\in\mathbb{C}^{1\times N}$, respectively, are modeled as
\begin{align}
{\bf h}_{\rm BI}=\sqrt{\zeta_{\rm BI}}{\bf g}_{\rm BI},~~ {\bf h}^H_{\rm IU}=\sqrt{\zeta_{\rm IU}}{\bf g}_{\rm IU}^H,
\end{align}
where $\zeta_{\rm BI}\triangleq\epsilon d_{\rm BI}^{-\alpha_{\rm BI}}$ and $\zeta_{\rm IU}\triangleq\epsilon d_{\rm IU}^{-\alpha_{\rm IU}}$ denotes the BS$\to$IRS and IRS$\to$UE link path-loss, respectively, with $d_{\rm BI}$ ($d_{\rm IU}$) being  the link distance and $\alpha_{\rm BI} (\alpha_{\rm IU})$ being the path-loss exponent. Moreover, ${\bf g}_{\rm BI}$ (${\bf g}_{\rm IU}^H$) denotes the corresponding small-scale fading channel, where $| g_{{\rm BI}, n}|$ and  $|g_{{\rm IU}, n}|$ , $n\in\mathcal{N}\triangleq\{1, \cdots, N\}$, both follow the Nakagami-$m$ distribution with parameters $m_{\rm BI}$ and $m_{\rm IU}$, respectively. Note that the Nakagami-$m$ fading is more general than the Rayleigh fading and can facilitate the performance analysis in the sequel. 

For the  active IRS, let ${\bf \Theta}\triangleq{\bf A}{\bf \Phi}\in\mathbb{C}^{N\times N}$ denotes its reflection matrix, where ${\bf \Phi}\triangleq{\rm diag}(e^{j\phi_{1}}, \cdots, e^{j\phi_{N}})$ denotes the IRS phase shift matrix with $\phi_{n}$ being the phase shift at each element $n\in\mathcal{N}$, and ${\bf A}\triangleq {\rm diag}(a_1, \cdots, a_N)$ denotes the active-IRS amplification matrix with $a_{n}$ being the amplification factor of each element $n$. Moreover, different from the passive IRS,  the active IRS incurs non-negligible thermal noise at all reflecting elements, which is denoted by $\boldsymbol{n}_{\rm F}\in\mathbb{C}^{N\times 1}$ and assumed following the distribution  
${\bf n}_{\rm F}\sim {\mathcal{CN}}({\bf 0}_{N}, \sigma^2_{\rm F}{\bf I}_{N})$ with $\sigma^2_{\rm F}$ denoting the amplification noise power.

Based on the above model, the received signal at the UE, denoted by $y$, can be modeled as follows. 
First, if the BS-UE distance is smaller than $L_{\rm in}$, we have 
\vspace{-0.2cm}
\begin{align}
y_1= h_{\rm BU}  x+ n_0,
\end{align} 
where $x$ is the transmitted signal with power $P_{\rm t}$ and $n_0$ denotes the received Gaussian noise at the user with power $\sigma^2$. Then the received SNR is given by
\vspace{-0.2cm}
\begin{equation}
{\rm SNR}_1=\frac{P_{\rm t} |h_{\rm BU}|^2}{\sigma^2}.
\end{equation}
Otherwise, the UE is associated with its nearest active IRS. For ease of analysis, we assume that the direct link is negligible due to the long distance and more severe blockage\footnote{It is noted that the proposed method can be used to obtain the result with the direct link taken into account, while the analysis is much more complicated yet without providing new insights.}. As such, the received signal is given by
\vspace{-0.2cm}
\begin{equation}
y_2 = {\bf h}_{\rm IU}^H {\bf A}{\bf \Phi}( {\bf h}_{{\rm BI}} x + {\bf n}_{\rm F}) + n_0.
\end{equation}
To maximize the UE's achievable rate, it is necessary to jointly design the reflection amplitude and phase shift of the active IRS, which is more complicated to the passive-IRS case that only involves the phase optimization. To this end, we first obtain its received SNR as below\footnote{The proposed method in this work can be extended to analyze the network performance in a more general multi-cell active-IRS aided wireless systems, where the inter-cell interference can be analyzed by a similar method for the amplification noise, which is more complicated and thus left for future work.}
\vspace{-0.2cm}
\begin{align}
{\rm SNR}_2=\frac{ P_{\rm t}| {\bf h}_{{\rm I}{\rm U}}^H {\bf A}{\bf \Phi} {\bf h}_{{\rm BI}} |^2 }{\Vert {\bf h}_{{\rm I} {\rm U}}^H {\bf A}{\bf \Phi} \Vert^2 \sigma_{\rm F} ^2 + \sigma^2}.\label{Eq:actSNR}
\end{align}
Next, due to the limited amplification power of the active IRS, the amplification matrix should satisfy the following constraint: $ P_{\rm t} \Vert {\bf A}{\bf \Phi} {\bf h}_{{\rm BI}}\Vert^2+\sigma_{\rm F}^2 \Vert  {\bf A}{\bf \Phi} {\bf I}_{N}\Vert^2 \le P_{\rm F}$, where $P_{\rm F}$ is the maximum amplification power of the active IRS.  Note that the active IRS amplifies both the received signal and non-negligible thermal noise at the reflecting elements. To simplify the analysis and facilitate practical implementation, we assume that all reflecting elements adopt a common amplification factor of $A$. As such, by following the similar reflection design method in \cite{you2021wireless}, the optimal active-IRS reflection design can be obtained as \footnote{The instantaneous CSI can be obtained by deploying sensors and then the active IRSs will feedback channel information to its connected BS for coordination.}
\begin{align}
[{\bf \Phi}^*]_n= &~ e^{j(-\angle{[{\bf h}_{{\rm I} {\rm U}}^H]_n}-\angle{[{\bf h}_{{\rm BI}}]_n})},\forall n,\label{Eq:beam} \\	(A^*)^2= &~\frac{P_{\rm F}}{P_{\rm t}\Vert{\bf h}_{{\rm BI}}\Vert^2+N\sigma_{\rm F}^{2}}=\frac{P_{\rm F}}{P_{\rm t} \zeta_{\rm BI}\Vert{\bf g}_{{\rm BI}}\Vert^2+N\sigma_{\rm F}^{2}}.\label{Eq:A}
\end{align}
Substituting \eqref{Eq:beam} and \eqref{Eq:A} into \eqref{Eq:actSNR} yields
\begin{align}
{\rm SNR}_2&
=\frac{ P_{\rm t}({A^*}|h_{\rm BIU}|)^2 }{(A^*)^2\Vert {\bf h}_{{\rm I} {\rm U}}^H \Vert^2 \sigma_{\rm F} ^2 + \sigma^2},
\end{align}
where 
$h_{\rm BIU}\triangleq\sum_{n=1}^N |{h}_{{\rm IU},n}| |{h}_{{\rm BI},n}|.$

\vspace{-0.3cm}
\subsection{Performance Metric}

Let $q({\rm SNR})$ denote a general function of the received {\text SNR}
which specifies the considered performance metric of the active-IRS aided wireless system. For example, when $q({\rm SNR})=\log_2(1+{\rm SNR})$, it represents the achievable rate in bits/second/Hz, while it represents the average and higher moments of SNR when $q({\rm SNR})={\rm SNR}^{\ell}$ with $\ell$ denoting the moment order. Further, we define $C={\mathbb{ E}_{\rm h}}[q({\rm SNR})]$ as the performance metric of the typical user averaged over channel fading, and $\bar{C}={\mathbb{ E}}[C]$ as the performance metric averaged over the distributions of all IRS and UE random locations. 

\section{Performance Analysis}

This section presents the channel power statistics for the involved channels and characterizes the communication performance using a mixture Gamma distribution approximation.


\subsection{Channel Power Statistics}
To facilitate the performance analysis, we first obtain the channel power statistics of the direct and cascaded channel under the Nakagami-$m$ distribution. Note that different from the existing works on IRS system performance analysis that usually adopt the moment matching methods, which achieve analytical tractability at the cost of degraded approximation accuracy, we utilize a \emph{mixture Gamma distribution} approximation to achieve more accurate and tractable results. As shown in \cite{chun2017comprehensive} and the references therein, the mixture Gamma approximation method provides higher accuracy than the moment matching methods.

First, we introduce the mixture Gamma distribution with its probability density function (PDF) and Laplace transform.
\begin{definition}{\rm Let $X$ denote a mixture Gamma distributed random variable. Its PDF and Laplace transform are characterized as follows.
\begin{equation}
    f_{X}(x) = \sum_{i=1}^{I}\varepsilon_{i}x^{\beta_{i}-1}e^{-\xi_{i}x},    \quad   \mathcal{L}_{X}(s) =  \sum_{i=1}^{I}  \varepsilon_i \frac{\Gamma(\beta_i)}{(\xi_i+s)^{\beta_i}},
\end{equation}  
where $\Gamma(\cdot)$ is the Gamma function and $\{\varepsilon_{i}, \beta_{i}, \xi_{i}\}$ are the parameters for each Gamma component.
}
\end{definition}

Next, we define 
\begin{align}
H_{\rm BU}&\triangleq |h_{\rm BU}|^2=\epsilon d_{\rm BU}^{-\alpha_{\rm BU}} |g_{\rm BU}|^2, \label{Eq:HBU}\\
H_{\rm BIU}&\triangleq ({A^*} |h_{\rm BIU}|)^2=(A^*)^2\l(\sum_{n=1}^N |{h}_{{\rm IU},n}| |{h}_{{\rm BI},n}|\r)^2\nn . 
\end{align} 
As $| g_{{\rm BU}}|$, $|g_{{\rm BI}, n}|$ and $|g_{{\rm IU},n}|$ follow the Nakagami-$m$ distribution, it can be easily shown that $H_{\rm BU}$ is  Gamma distributed and $H_{\rm BIU}$ can be accurately approximated by a mixture Gamma distribution, which is parameterized as follows.\footnote{For ease of notation,  we simply use $\alpha$ to represent the path-loss exponent in the sequel for each individual link without causing confusion. }

\begin{lemma}\label{Lem:H_BU}\emph{$H_{\rm BU}$ in \eqref{Eq:HBU} follows the Gamma distribution, which can also be molded as a mixture Gamma distribution, parameterized by  $I=1$, and 
\begin{small}
\begin{align}
\varepsilon_{{\rm BU}} = \frac{({d_{\rm BU}^{\alpha }m_{\rm BU}})^{m_{\rm BU}}}{\epsilon^{m_{\rm BU}}\Gamma(m_{\rm BU})}, \quad
\beta_{{\rm BU}} = m_{\rm BU}, \quad
\xi_{{\rm BU}} = m_{\rm BU}\frac{d_{\rm BU}^{\alpha}}{\epsilon}.
\end{align} 
\end{small}} 
\end{lemma}

\begin{lemma}\label{Lem:HBIU}\emph{$H_{\rm BIU}$ in \eqref{Eq:HBU} can be approximated by the mixture Gamma distribution, parameterized by  
\begin{small}
\begin{align}\label{eq: MG_para_BIU}
\begin{split}
    &~ \varepsilon_{{\rm BIU},i} = \frac{(m_{\rm BI}m_{\rm IU})^{m_{\rm BI}}w_{i}t_{i}^{m_{\rm IU}-m_{\rm BI}-1}}{\Gamma(m_{\rm BI})\Gamma(m_{\rm IU})}\left({\frac{W}{(A^{*})^2N^2}}\right)^{m_{\rm BI}}, \\ &~
    \beta_{{\rm BIU},i} = m_{\rm BI}, \quad
    \xi_{{\rm BIU},i} = \frac{m_{\rm BI}m_{\rm IU}}{t_{i}}{\frac{W}{(A^{*})^2N^2}},
\end{split}
\end{align} 
\end{small}where $t_{i}$ is the $i$-th zero of Laguerre polynomials and $w_{i}=\frac{2^{n-1}n!\sqrt{\pi}}{n^2[H_{n-1}(t_i)]^2}$ is the relative $i$-th weight factor, $W= \frac{{d^{\alpha}_{\rm BI}}{d^{\alpha}_{\rm IU}}}{\epsilon^2}$, and $I$ is set as $I=20$ to guarantee sufficient accuracy (i.e., approximation error less than $10^{-5}$).} 
\end{lemma}

Sketch of proof: This can be proved by simplifying the Meijer-G function and utilizing the Laguerre polynomials \cite{li2022analysis}.

\vspace{-0.3cm}
\subsection{General Results of Performance Metrics}

We first adopt a key lemma below that characterizes the system performance by the function of received SNR \cite{hamdi2007useful}. 

\begin{lemma}\emph{
If $X_1$ follows the mixture Gamma distribution parameterized by $\{ \varepsilon_{i}, \beta_{i}, \xi_{i} \}$, and $X_{2}$ is independent of $X_{1}$, then we have
\vspace{-0.2cm}
\begin{small}
\begin{equation}
    \mathbb{E} \left[q\left(\frac{X_1}{X_2+b}\right)\Big\vert X_{2}\right] =\sum_{i=0}^{I} \varepsilon_{i} \Gamma(\beta_{i}) {\xi_{i}}^{-\beta_{i}}  \int_{0}^{\infty} {q_{\beta_{i}}(z) e^{- {\xi_{i}} z(x_2+b)}}  \mathrm{d}z ,
\end{equation}
\end{small}where $z\triangleq\frac{x_1}{x_2+b}$ represents the SNR, $q_{\beta_{i}}(z) = \frac{1}{\Gamma(\beta_{i})}\frac{\mathrm{d}^{\beta_{i}}}{\mathrm{d}z^{\beta_{i}}}g(z) $. For the performance of achievable rate, we have $q(z)=\ln(1+z)$ and thus $q_{\beta_{i}}(z) =\frac{1}{z} -\frac{1}{z(1+z)^{\beta_{i}}}$; while for the SNR moments, we have $q(z)=z^{\ell}$ and thus $q_{\beta_{i}}(z) = \frac{\Gamma(\beta_{i}+l)}{\Gamma(\ell)\Gamma(\beta_{i})}z^{\ell-1}$. 
}
\end{lemma}

Sketch of proof: This is extended from the system analysis of Nakagami-$m$ distributed traditional wireless networks \cite{hamdi2007useful}.

\subsubsection{Analysis for Conditional SNR Moments}
First, we characterize the conditional SNR moments given the locations of the user and IRS for the following two cases.

{\bf Case 1:} The user is inside the BS-coverage region ($d_{\rm BU}<L_{\rm in}$), and the received signal power follows the Gamma distribution. Then, its SNR moments conditioned on the locations of the user can be obtained as follows by using  Lemma~\ref{Lem:H_BU},
\begin{small}
\begin{equation} \label{eq: SNR_m_d}
    \begin{split}
        C_{1}^{\rm(SNR)} = & \mathbb{E}_{\rm h} [{\rm SNR}_1^{\ell}] = 
        \frac{\Gamma(m_{\rm BU}+\ell) }{\Gamma(m_{\rm BU})} \left( \frac{ m_{\rm BU} d_{\rm BU}^{\alpha} \delta^{2}}{\epsilon P_{\rm t}} \right)^{-\ell}.
    \end{split}
\end{equation}  
\end{small}

{\bf Case 2:} The user is outside the BS-coverage region ($d_{\rm BU}\ge L_{\rm in}$), and it is assisted by the active IRS. Since the received signal power follows the mixture Gamma distribution as shown in Lemma~\ref{Lem:HBIU}, its conditional SNR moments can be obtained as below by following the similar method in \cite{li2022analysis}
\vspace{-0.2cm}
\begin{small}
\begin{equation}  \label{eq: SNR_m_ir}
    \begin{split}
        C_{2}^{\rm(SNR)} = &~ \mathbb{E}_{\rm h} [{\rm SNR}_2^{\ell}] =  \sum_{i=0}^{I} \varepsilon_{{\rm BIU},i} \Gamma(\beta_{{\rm BIU},i} ) \xi_{{\rm BIU},i}^{-\beta_{{\rm BIU},i} } \\ \!\!\! \cdot & \int_{0}^{\infty} { \frac{\Gamma(\beta_{{\rm BIU},i} +\ell) }{\Gamma(\beta_{{\rm BIU},i} )\Gamma(l)} z^{\ell-1} e^{-z\xi_{{\rm BIU},i}  \frac{\delta^2}{P_{\rm t}}} } \mathcal{L}_{\rm N_{\rm F}}( z)  \mathrm{d}z.
    \end{split} 
\end{equation}
\end{small}Note that $ \mathcal{L}_{\rm N_{\rm F}}( z) $ is the Laplace transform of the received thermal noise generated by active IRS, which is given by
\begin{small}
\begin{align} \label{eq: laplaceThermalN}
     \mathcal{L}_{\rm N_{\rm F}}(z)  &=    \mathbb{E}_{\rm h}\left[\exp\left(-z\frac{(A^{*})^2N\delta_{\rm F}^{2}{\xi_{{\rm BIU},i}}}{P_{\rm t}}{H_{\rm IU}}\right) \right]\nn 
     \\ &\overset{(a)}{\approx}\frac{m_{\rm IU}^{-m_{\rm IU}}}{\left(m_{\rm IU}+z\frac{(\overline{A^{*}})^2N\delta_{\rm F}^{2}{\xi_{{\rm BIU},i}}}{P_{\rm t}}\right)^{m_{\rm IU}}},
\end{align}
\end{small}where $\overline{(A^{*})^2}=\frac{P_{\rm F}}{N(P_{\rm t}\epsilon d_{\rm BI}^{-\alpha}+\delta_{\rm F}^{2})}$, and $(a)$ holds since it can be easily shown that $H_{\rm IU}\triangleq \Vert{\bf h}_{\rm IU}\Vert^2= \epsilon d_{\rm IU}^{-\alpha_{\rm IU}} \sum_{n=1}^N|g_{{\rm IU},n}|^2$ follows the Gamma distribution and its parameters can be obtained by following the similar method in  Lemma~\ref{Lem:H_BU}. 

Based on \eqref{eq: SNR_m_ir} and \eqref{eq: laplaceThermalN}, the mean SNR is obtained below.
\begin{lemma}\label{Proposition:SNR}\emph{When the UE is located outside the BS-coverage region, the mean SNR of active IRS aided communication is given by 
\vspace{-0.2cm}
\begin{small}
\begin{equation} \label{eq:meanSNROrigin}
    \mathbb{E}_{\rm h}[{\rm SNR}_2] = \sum_{i=0}^{I} \frac{w_{i}t_{i}^{2m_{\rm IU}-1}}{\Gamma(m_{\rm IU})}m_{\rm BI}^{1-m_{\rm IU}}\left( \frac{P_{\rm t}N}{m_{\rm IU}^{2}\delta_{\rm F}^{2}W} \right)^{m_{\rm IU}}\varphi(z),
\end{equation}
\end{small}where $\varphi(z) = \int_{0}^{\infty}e^{-z\frac{m_{\rm BI}m_{\rm IU}W\delta^2}{t_{i}\eta N P_{\rm t}}}{\left( z + \frac{NP_{\rm t}t_{i}}{\delta_{\rm F}^{2}m_{\rm BI}W} \right)^{-m_{\rm IU}}} {\rm d} z$ and $\eta = \frac{P_{\rm F}}{P_{\rm t}\epsilon d_{\rm BI}^{-\alpha}+\delta_{\rm F}^2}$. }
\end{lemma}
\begin{proposition}\emph{The closed-form expression for the mean SNR can be obtained in \eqref{eq:meanSNRclosedF} by substituting \cite[eq.3.353.2]{gradshteyn2014table}, where $\Upsilon=\left(\frac{P_{\rm t} t_{i}}{W\delta_{\rm F}^{2} m_{\rm BI}m_{\rm IU}^{2}}\right)^{m_{\rm IU}}\frac{m_{\rm BI}{w_{i}t_{i}^{m_{\rm IU}-1}}}{\Gamma(m_{\rm IU})(m_{\rm IU}-1)!}$, and $E_{i}(x) = \int_{-x}^{\infty} \frac{e^{-t}}{t}{\rm d}t$ is the exponential integral function. } 
\begin{figure*}[t]
\begin{small}
\begin{equation} \label{eq:meanSNRclosedF}
    \begin{split}
    \mathbb{E}_{\rm h}[{\rm SNR}_2] = &~ \sum_{i=1}^{I} N^{m_{\rm IU}} \Upsilon  \left[ \sum_{k=1}^{m_{\rm IU}-1} (k-1)! \left( -\frac{m_{\rm BI}m_{\rm IU}W\delta^{2}}{t_{i}N\eta P_{\rm t}} \right)^{m_{\rm IU}-1-k} \left( \frac{t_{i}N P_{\rm t}}{m_{\rm BI}W\delta_{\rm F}^{2}} \right)^{k}   - \left( -\frac{m_{\rm BI}m_{\rm IU}W\delta^{2}}{t_{i}N\eta P_{\rm t}} \right)^{m_{\rm IU}-1} e^{\frac{m_{\rm IU}\delta^2}{\eta\delta_{\rm F}^2}} E_{i}\left(-\frac{m_{\rm IU}\delta^2}{\eta\delta_{\rm F}^2}\right)  \right],  
    \end{split}
\end{equation}
\end{small}
\hrulefill
\end{figure*} 
\end{proposition}

Specifically, for $m_{\rm IU}=1$ corresponding to the Rayleigh fading case, we have the following result.

\begin{figure} [h]
    \centering
    \includegraphics[height=2.5cm,width=6cm]{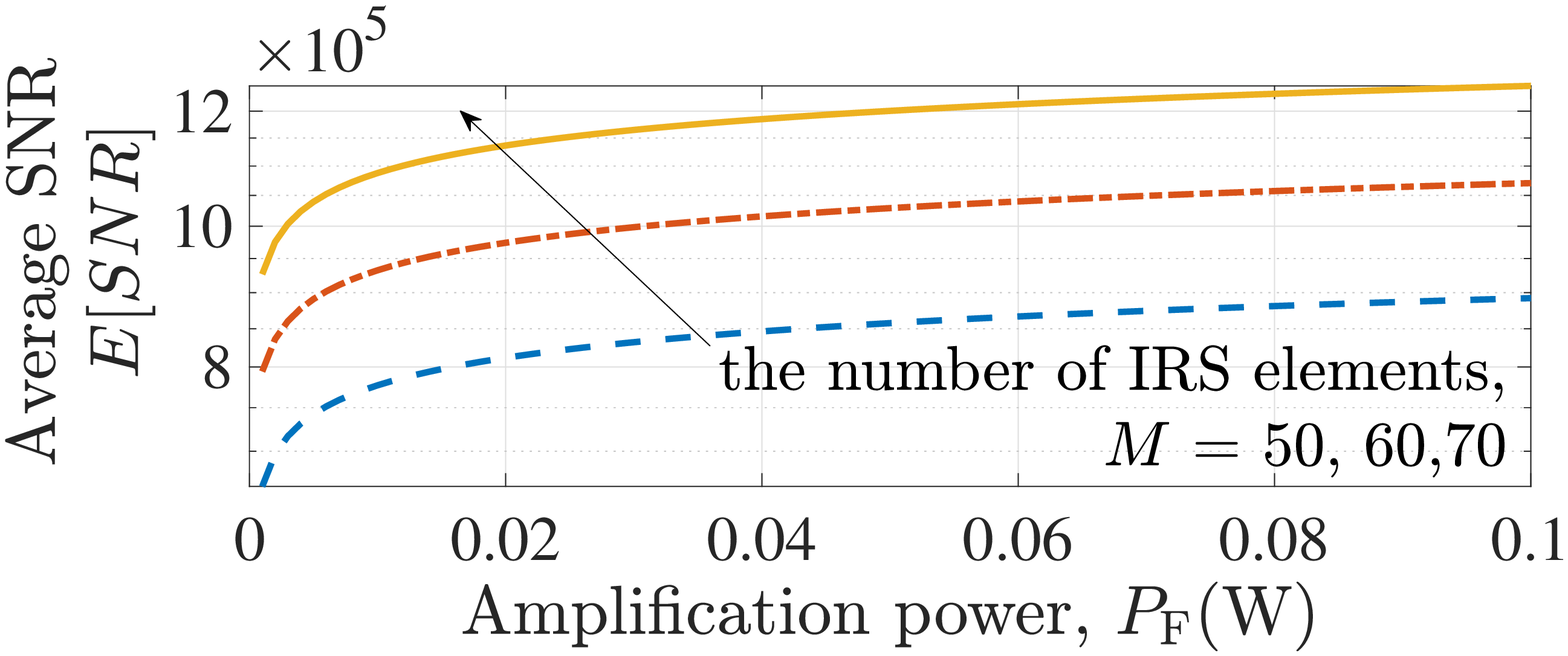}
    \caption{Mean SNR under different amplification power.}
    \label{fig:PF}
\end{figure}

\begin{corollary}\emph{When $m_{\rm IU}=1$, the mean SNR of the active-IRS aided wireless network for the case of $d_{\rm BU}\ge L_{\rm in}$ is given by 
\begin{align} \label{eq:Pf_effect}
        \mathbb{E}_{\rm h}[{\rm SNR}_2] = \frac{N P_{\rm t}}{W\delta_{\rm F}^{2}} \exp\left({\frac{\Psi}{P_{\rm F}}}\right) E_{1}\left(\frac{\Psi}{P_{\rm F}}\right),
\end{align}
where $\Psi\triangleq\frac{\delta^{2}(P_{\rm t}\epsilon d_{\rm BI}^{-\alpha}+\delta_{\rm F}^{2})}{\delta_{\rm F}^{2}}$.
}
\end{corollary}
Although it is still difficult to characterize the effect of the amplification power on the mean SNR from \eqref{eq:Pf_effect}, one can observe from the numerical result, Fig.~\ref{fig:PF}, that the mean SNR first increases quickly and then slows down when the amplification power increases.

\begin{remark}\emph{One can observe from \eqref{eq:Pf_effect} that the mean SNR of the active IRS linearly increases with the number of IRS elements, i.e., $ \mathbb{E}_{\rm h}[{\rm SNR}_2]\sim \mathcal{O}(N)$, which is consistent with \cite{you2021wireless}.
}
\end{remark}

For comparison, we also analyze the mean SNR of the passive-IRS case. Following the similar procedures in Lemma~\ref{Proposition:SNR}, the mean SNR of the passive-IRS case (denoted by $\widetilde{\rm SNR}$) is obtained as
\vspace{-0.2cm}
\begin{align} \label{eq: mean_SNR_specialC_p}
    &~ \mathbb{E}_{\rm h}[\widetilde{\rm SNR}] = N^2 \sum_{i=0}^{I} \frac{w_{i}t_{i}^{m_{\rm IU}}P_{\rm t}}{\Gamma(m_{\rm IU}+1)\delta^{2}W},
\end{align}
which shows that the mean SNR of the passive-IRS assisted system increases with $N$ in the order of $\mathcal{O}(N^2)$. It is worth mentioning that although the passive IRS has a larger power scaling order than the active IRS, i.e.,  $\mathcal{O}(N^2)$ versus $\mathcal{O}(N)$, the actual SNR for small or modest $N$ depends on the amplification factor of each IRS elements.


\subsubsection{Analysis for Conditional Achievable Rate}

Following the similar derivation of conditional SNR moments, the conditional achievable rate for Case 1 and Case 2 can be obtained as \eqref{eq:C_AR_1} and \eqref{eq:C_AR_2}, respectively.
\vspace{-0.1cm}
\begin{small}
\begin{equation}
    C_{1}^{\rm(AR)} =  \log_{2}e\int_{0}^{\infty} { \frac{1}{z}\left( 1-\frac{1}{(1+z)^{m_{\rm BU}}} \right) e^{-\frac{m_{\rm BU}d_{\rm BU}^{\alpha} \delta^2}{\epsilon P_{t}} z}} \mathrm{d}z, \label{eq:C_AR_1}
\end{equation}
\begin{equation}
    \begin{split}
        C_{2}^{\rm(AR)} =&~ \log_{2}e \sum_{i=0}^{I}  \varepsilon_{{\rm BIU},i} \Gamma(\beta_{{\rm BIU},i}) {\xi_{{\rm BIU},i}}^{-\beta_{{\rm BIU},i}} \\ &~ \cdot  \int_{0}^{\infty} { \frac{1}{z}\left( 1-\frac{1}{(1+z)^{\beta_{{\rm BIU},i}}} \right) e^{-\frac{\delta^2}{P_{t}}\xi_{{\rm BIU},i} z}} \mathcal{L}_{\rm N_{\rm F}}(z) \mathrm{d}z . \label{eq:C_AR_2}
    \end{split} 
\end{equation}
\end{small}


\subsubsection{Analysis for Average Performance Analysis}
Based on the above, we characterize the average SNR and spatial throughput over the distributions of all UEs and IRSs based on the conditional SNR moments and achievable rate in \eqref{eq: SNR_m_d}, \eqref{eq: SNR_m_ir}, \eqref{eq:C_AR_1}, and \eqref{eq:C_AR_2}. Specifically, the cell can be divided into three areas: the part directly served by BS  with $d_{\rm BU}\leq L_{\rm in}$, the part served by the active IRS and located within the ring with $L_{\rm in}<d_{\rm BU}< L_{\rm out}$, and the part served by the active IRS and located outside the ring with $d_{\rm BU}\geq L_{\rm out}$. As such, the average system performance can be obtained as $
    \mathbb{E} [{\rm C}]  = \frac{S_{1}}{S_{\rm t}}\overline{C_{1}}+ \frac{S_{2}}{S_{\rm t}}\overline{C_{2}}+ \frac{S_{3}}{S_{\rm t}}\overline{C_{3}},$
where $\overline{C_{1}}=\mathbb{E}[C_{1}]\vert_{d_{\rm BU}=0}^{L_{\rm in}}$, $\overline{C_{2}}=\mathbb{E}[C_{2}]\vert_{d_{\rm BU}=L_{\rm in}}^{L_{\rm out}}$, and $\overline{C_{3}}=\mathbb{E}[C_{2}]\vert_{d_{\rm BU}=L_{\rm out}}^{L}$ denote the average system performance in the corresponding distance range, respectively, which are weighted by their areas (thanks to uniform UE distribution) given by $S_{\rm t}=\pi L^{2}$, $S_{1}=\pi L_{\rm in}^{2}$, $S_{2}= \pi( L_{\rm out}^{2} - L_{\rm in}^{2})$, and $S_{3}= \pi( L^{2} - L_{\rm out}^{2})$. When $d_{\rm BU}\leq L_{\rm in}$, the PDF of $d_{\rm BU}$ is $f_{d_{\rm BU}}(d_{\rm BU}) \triangleq \frac{2\pi d_{\rm BU}}{S_{1}}$ \cite{lyu2020spatial}. When $L_{\rm in}\leq d_{\rm BU}\leq L_{\rm out}$, the PDF of $d_{\rm BU}$ is $f_{d_{\rm BU}}(d_{\rm BU}) \triangleq \frac{2\pi d_{\rm BU}}{S_{2}}$. To facilitate the analysis, we assume that $d_{\rm BI}\thickapprox d_{\rm BU} $ \cite{lyu2020spatial}. The PDF of the distance from a typical UE to its nearest IRS is $f_{\rm d_{\rm IU}}(d_{\rm IU}) \triangleq  2\pi\lambda_{\rm I}d_{\rm IU} e^{-\lambda_{\rm I}\pi d_{\rm IU}^{2}}$, where $\lambda_{\rm I}=M/S_{\rm 2}$. When $d_{\rm BU} \geq L_{\rm out}$, the PDF of $d_{\rm BU}$ is $f_{d_{\rm BU}}(d_{\rm BU}) \triangleq \frac{2\pi d_{\rm BU}}{S_{3}}$. In addition, we assume that $d_{\rm BI}\thickapprox L_{\rm out}$ and $d_{\rm IU}\thickapprox d_{\rm BU} - L_{\rm out}$. As such, the average performance is given by \eqref{eq:averagePerf}.
\begin{figure*}
\begin{equation}\label{eq:averagePerf}
    \mathbb{E}[C] = \frac{2\pi}{S_{\rm t}} \left[ \int_{d_{\rm BU}=0}^{L_{\rm in}} C_{1}d_{\rm BU}\mathrm{d}d_{\rm BU} + \int_{d_{\rm BU}=L_{\rm in}}^{L_{\rm out}} \int_{d_{\rm IU}=0}^{L} C_{2}d_{\rm BU}f_{d_{\rm IU}}(d_{\rm IU}) \mathrm{d}d_{\rm BU}\mathrm{d}d_{\rm IU} + \int_{d_{\rm BU}=L_{\rm out}}^{L}C_{2}d_{\rm BU} \mathrm{d}d_{\rm BU}\right].
\end{equation}
\hrulefill
\end{figure*}

\vspace{-0.25cm}
\section{Numerical results} \label{Section: numerical results}

\begin{figure}[!t]
	\centering
	\subfigure[Active IRS deployment.]{
		\begin{minipage}[b]{0.22\textwidth}
			\includegraphics[height=3.4cm,width=4.4cm]{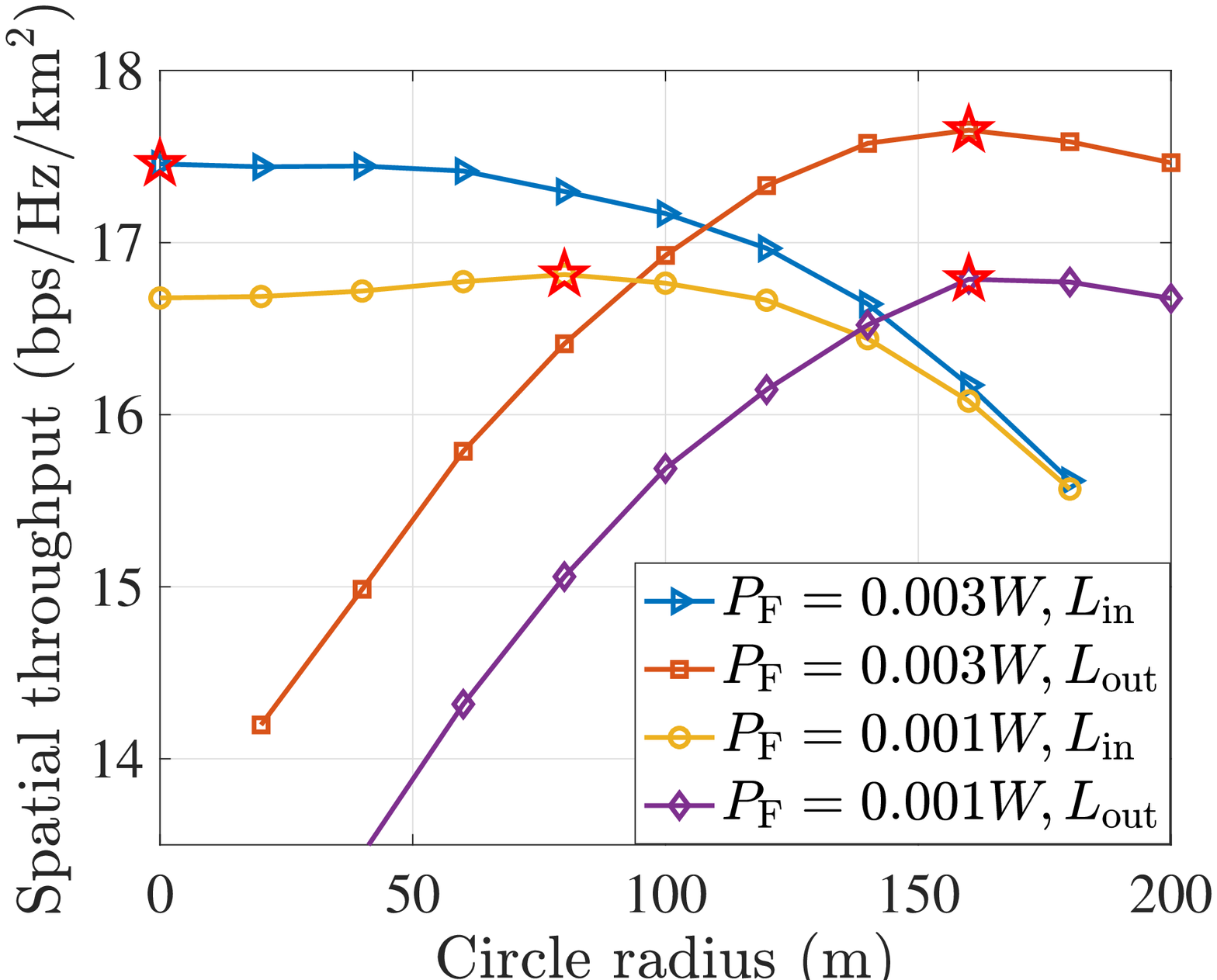}
		\end{minipage}
		\label{fig:innerAndOuterCircleRadius}
	}
    \subfigure[Association policy.]{
    	\begin{minipage}[b]{0.22\textwidth}
   		    \includegraphics[height=3.4cm,width=4.4cm]{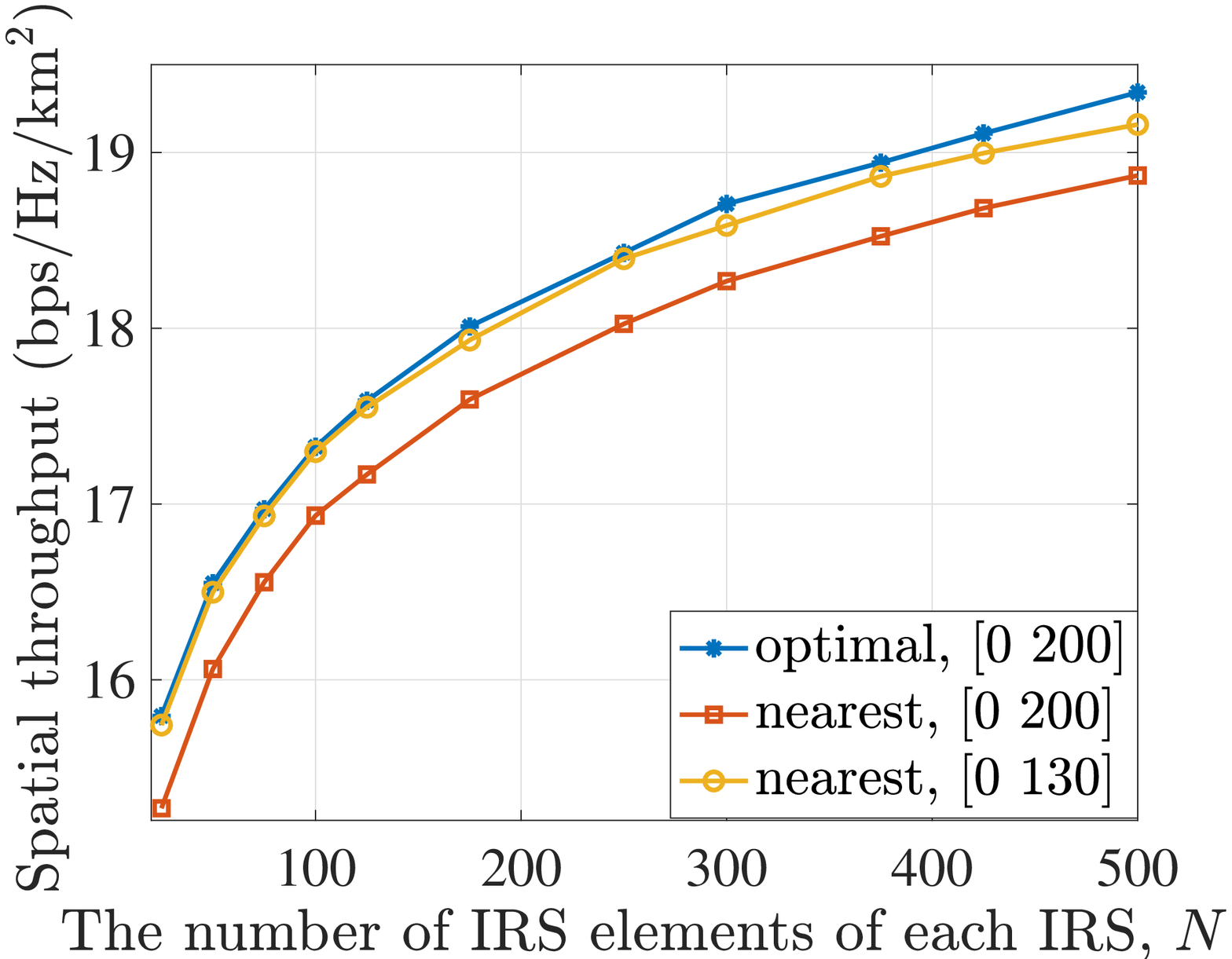}
    	\end{minipage}
	    \label{fig:associationP}
    }
	\caption{Effectiveness of the proposed IRS deployment and association policy.}
	\label{fig:setup}
    \subfigure[Average SNR comparison of passive and active IRS.]{
		\begin{minipage}[b]{0.22\textwidth}
			\includegraphics[height=3.4cm,width=4.4cm]{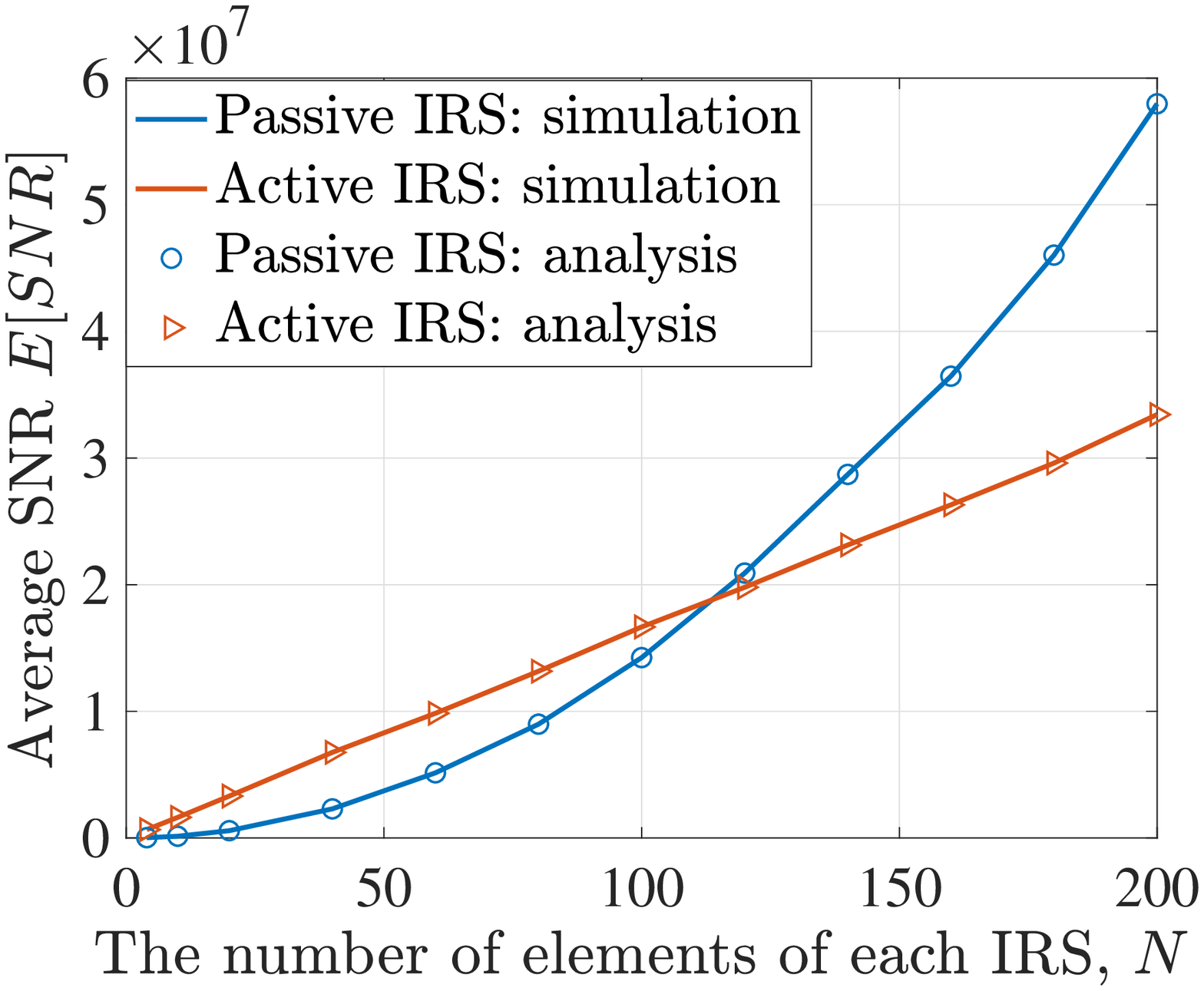}
		\end{minipage}
		\label{fig:SNR_ave}
	}
    \subfigure[Spatial throughput versus number of IRSs given fixed number of total IRS elements.]{
    	\begin{minipage}[b]{0.22\textwidth}
   		    \includegraphics[height=3.4cm,width=4.4cm]{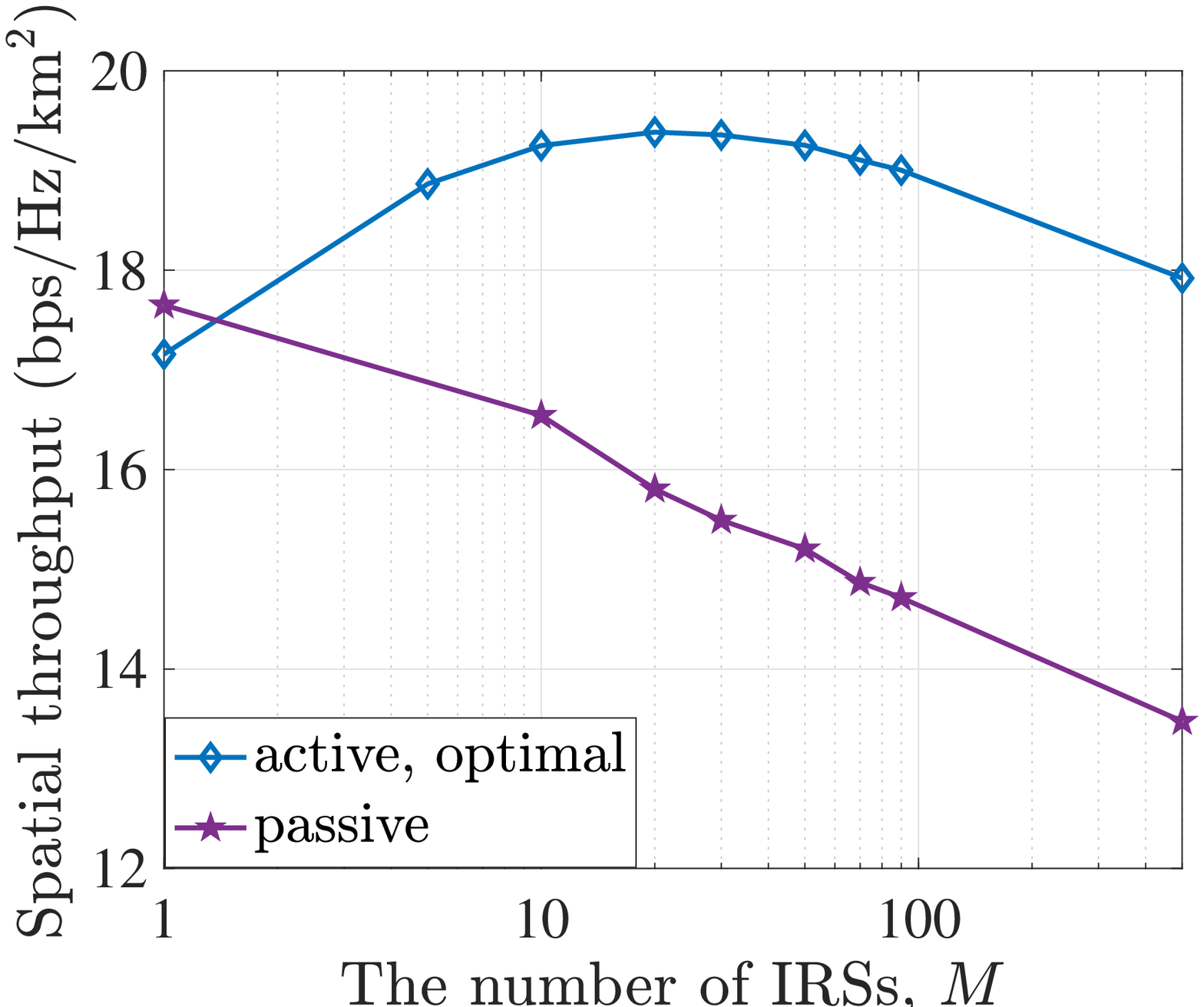}
    	\end{minipage}
	     \label{fig:fixTotNumOfIRSele_activepassive}}
	\caption{Performance comparison with passive IRS.}
	\label{fig:perf}
\end{figure}

In this section, Monte-Carlo (MC) simulation with $10^{6} $ iterations are carried out in MATLAB 
to verify our analysis and compare the performance between the active- and passive-IRS aided system. The simulation setup is as follows, if not specified otherwise. The path loss exponent is $\alpha=3$, the transmit power of BS is $P_{\rm t} = 1$ W,  $L=200$ m, $\delta^2 = -80$ dBm, and $\delta_{\rm F}^2 = -70$ dBm \cite{you2021wireless}.

\subsubsection{Effectiveness of the Proposed IRS Deployment and Association Policy}

In Fig.~\ref{fig:innerAndOuterCircleRadius}, we show the effects of the inner and outer range on the performance of the spatial throughput. 
It is observed that there exist optimal values for both the inner and outer ranges, which vary with the amplification power of the active IRS, i.e., $P_{{\rm F}}$. Specifically, when $P_{{\rm F}}$ is small, the active IRS should be deployed sufficiently far from the BS such that it can amplify the signal, which is consistent with the placement optimization of the active IRS in the link-level analysis \cite{you2021wireless}. 
In Fig.~\ref{fig:associationP}, we show the effectiveness of the proposed user association policy for the active-IRS aided system. It is observed that the considered nearest association policy with $L_{\rm out}=130$ m and a relatively small number of IRS elements achieves very close performance with the optimal one that associates each user to its best active IRS. Intuitively, the ring deployment strategy guaranteed that the nearest active IRS is located close to the optimal location. 


\subsubsection{Performance Comparison with Passive IRS}
In Fig.~\ref{fig:SNR_ave}, we compare the average SNR between the passive and active IRS, given link distances. It is observed that the average SNR of the active IRS increases linearly with $N$, while that of the passive IRS increases much faster when $N$ is sufficiently large due to a higher power scaling order. Besides, the active IRS outperforms the passive IRS when $N$ is small and vice versa. This is expected since for small  $N$, the passive IRS yields limited power gain while the active IRS provides additional power amplification gain. 
Fig.~\ref{fig:fixTotNumOfIRSele_activepassive} shows the spatial throughput of the active- and passive-IRS aided system versus the number of IRSs, given a fixed budget on the total number of IRS elements and power consumption. Several interesting observations are made as follows. First, for the passive-IRS case, it is beneficial to assemble all reflecting elements into one single IRS (i.e., centralized deployment strategy). In contrast, there generally exists an optimal active-IRS density (i.e., the number of IRS in the network) for the active-IRS case. This can be explained by the fact that the maximum amplification factor provided by the active IRS is constrained by not only its amplification power but also the number of IRS elements.


\vspace{-0.25cm}

\section{Conclusions}

In this letter, we characterize the communication performance of a single-cell active-IRS aided wireless network. To this end, we first propose a \emph{customized} IRS deployment strategy and then apply the  \emph{mixture Gamma distribution} approximation method to obtain a closed-form expression for the mean SNR at the user averaged over the Nakagami-$m$ channel fading.
Moreover, we numerically show that to maximize the spatial throughput, it is necessary to choose a proper active-IRS density given a fixed number of total reflecting elements, which significantly differs from the passive-IRS case for which a centralized IRS deployment scheme is optimal. Furthermore, the active-IRS aided network achieves higher throughput than the passive-IRS counterpart when the total number of reflecting elements is small. This work can be extended to account for network coordination, outdated channel state information, multiple reflections, and multi-cell networks.

\appendices

\bibliographystyle{IEEEtran}  
\bibliography{references}

\newpage

\end{document}